\documentclass[annals, final,3.5p,times]{elsarticle}

\usepackage{hyperref}
\usepackage{graphicx}
\usepackage{amssymb}
\usepackage{amsmath}
\usepackage{cases}
\usepackage{xcolor}

\newcommand{\bs}[1]{{\boldsymbol{#1}}}
\newcommand{\bk}{\bs{k}}
\newcommand{\br}{\bs{r}}

\begin{document}
\begin{frontmatter}

\title{Coherent multiple scattering of out-of-equilibrium interacting Bose gases}

\author{Nicolas Cherroret}
\ead{cherroret@lkb.upmc.fr}
\author{Thibault Scoquart}
\ead{scoquart@lkb.upmc.fr}
\author{Dominique  Delande}
\ead{delande.upmc.fr}
\address{Laboratoire Kastler Brossel, Sorbonne Universit\'e, CNRS, ENS-Universit\'e PSL, Coll\`ege de France; 4 Place Jussieu, 75004 Paris, France }

\begin{abstract}
We review recent theoretical and experimental progresses in the coherent multiple scattering of weakly interacting disordered Bose gases. These systems have allowed, in the recent years, a characterization of weak and strong localization phenomena in disorder at an unprecedented level of control. In this paper, we first discuss the main physical concepts and recent experimental achievements associated with a few emblematic ``mesoscopic'' effects in disorder like coherent back scattering, coherent forward scattering or mesoscopic echos, focusing on the context of out-of-equilibrium cold-atom setups. 
We then address the role of weak particle interactions and explain how, depending on their relative strength with respect to the disorder and on the time scales probed, they can give rise to a dephasing mechanism for weak localization, thermalize a non-equilibrium Bose gas or make it become a superfluid.
\end{abstract}


\end{frontmatter}

\section{Introduction}

\subsection{Coherent multiple scattering in solids}

In disordered conductors, the question of how interference in multiple scattering affects transport observables like the conductance has been central since the 70's \cite{Datta95}. When temperature is not low enough interference is usually negligible due to the too small electron coherence length $L_\phi$: the conductance faithfully obeys Ohm's law, which describes electron multiple scattering as a classical diffusive  process through the material's impurities. The conductance is also self-averaging: its value is independent of the specific configuration of impurities because the material behaves as a superposition of many incoherent sub-systems of size $L_\phi$.
The situation changes at low temperature, in the mesoscopic regime where $L_\phi$ exceeds the conductor size and electrons genuinely behave like coherent waves. 
The  conductance is no longer self-averaging, and deviations to Ohm's law appear \cite{Gorkov79}. These deviations, known as weak localization corrections, stem at leading-order from the enhanced probability for electrons to return to a point already explored, by constructive interference between reversed multiple scattering trajectories. Weak localization was observed in many systems, especially in thin films where a magnetic field can be used as a knob to turn off weak localization in a controlled way \cite{Bergmann84}. Its behavior in the presence of electron-electron interactions was also extensively studied \cite{Altshuler80, Altshuler82} and, nowadays, weak localization measurements 
still constitute a valuable tool to assess electronic coherence \cite{Niimi09, Capron13}.

In the 80's, several experiments also reported a vanishing of the electron conductivity in  the zero-temperature limit of strongly disordered conductors \cite{Hertel83, Paalanen83, Rosenbaum94, Katsumoto87}, attributed to Anderson  localization of electrons. In three dimensions, this phenomenon, discovered by Anderson in 1958 \cite{Anderson58}, manifests itself as a quantum phase transition: the electron spectrum displays a mobility edge, a critical energy above which the eigenstates of the disordered system are spatially extended, and below which they are exponentially localized. The activity on Anderson localization surged with the development of the celebrated scaling theory \cite{Abrahams79}, based on precursory works by Thouless \cite{thouless74}. Beyond providing an elegant description of Anderson localization in terms of scaling arguments, this approach also predicted, in addition to a mobility edge in three dimensions, the localization of all eigenstates in one-dimensional (1D) and two-dimensional (2D) disordered systems. Furthermore, the scaling theory made contact between Anderson localization and the weak localization corrections, describing the latter as a precursor of the former. Since then, several theoretical descriptions of the localization problem have been developed, from diagrammatic formalisms \cite{Berezinskii74}, random-matrix approaches \cite{Dorokhov82, Mello88}, field theories \cite{Efetov80}, to approximate self-consistent treatments \cite{Vollhardt80a}. Today,  a broad class of Anderson transitions has been identified beyond the historical ``Wigner-Dyson'' ensembles where only time-reversal and spin-rotation symmetries matter \cite{Altland97, Evers08}, and continue to be explored in a large spectrum of condensed-matter systems.

\subsection{Localization phenomena with cold atoms}

The first experimental studies on Anderson localization of atomic matter waves date back from the early 2000's. These experiments involve clouds of cold atoms evolving in far-detuned spatially disordered optical potentials usually produced from laser speckles or bichromatic lattices. 
As compared to conduction electrons in solids, cold-atom setups have the advantage of offering a great control of the disorder and the possibility to measure local observables. Atomic interactions can also be tuned to a large extent, and therefore studied in a systematic way \cite{Bloch08, Shapiro12, SanchezP10}. In contrast, electron interactions in solids are hardly controllable. Anderson localization of non-interacting atomic matter waves was first observed in one dimension \cite{Billy08, Roati08}. These experiments were accompanied by a number of theoretical works, \cite{SanchezP07, kuhn07, SergeyBEC07} to cite a few. In three dimensions, three experiments reported on Anderson localization of cold atoms as well \cite{Jendrzejewski12b, Kondov11, Semeghini15}, although the results of \cite{Kondov11} have been severely questioned \cite{Pasek17}. These  observations were all based on a fundamental signature of Anderson localization, the temporal freezing of a spreading wave packet. In these works, however, the properties of the Anderson transition were not characterized in detail, mostly due to the rather broad energy distribution of the atoms near the transition. This task was accomplished in a different system, the ``atomic kicked rotor'', in a series of experiments \cite{Chabe08, Lemarie10, Lopez12} (see also \cite{Manai15} for an observation of  Anderson localization in two dimensions). 
Unlike electronic systems, in atomic gases the observations of Anderson localization predated the detection of weak localization. The latter was first observed and characterized in 2012 via the (atomic) coherent backscattering effect \cite{Jendrzejewski12, Labeyrie12}, followed a few years later by an observation of a related effect, the mesoscopic echo \cite{Hainaut17},  in  an experimental realization of the atomic kicked rotor. Alongside, a recent experimental feat was the ability to selectively break the interference between time-reversed trajectories at play in weak localization by means of a controlled dephasing for cold atoms \cite{Muller15}. Similar in spirit, the application of an artificial gauge field was recently used to drive a localized atomic system from the orthogonal to the unitary symmetry class and to foster the emergence of a novel interference mechanism in disorder known as coherent forward scattering \cite{Hainaut18}. These ``spectroscopies'' of interference scattering sequences \cite{Micklitz15} recall the conductivity measurements methods used in mesoscopic physics, but push the experimental control to an unprecedented level. 

In contrast with conduction experiments, usually based on conductivity measurements that involve electrons in thermal equilibrium, many cold-atoms setups  probe the dynamics of a quantum  gas put \textit{out-of-equilibrium} after a ``quench'' \cite{Polkovnikov11, Altman12} (see, however, also \cite{Krinner13, Krinner15, Hartung08} for  stationary transport configurations more in the spirit of mesoscopic physics). In the presence of both disorder and interactions, the non-equilibrium dynamics of quantum gases is especially rich and  yet largely unexplored, interactions tending to thermalize the gas while disorder tending to localize it. This competition is the essence of the many-body localization problem, which manifests itself as a dynamical phase transition between an ergodic (thermal) phase and a many-body localized phase at strong enough disorder \cite{Gornyi05, Basko06, Nandkishore15, Alet18, Abanin19}, whose signatures were recently observed experimentally \cite{Schreiber15, Bordia16, Choi16}.

In this article, we review recent developments in the context of coherent multiple scattering of out-of-equilibrium Bose gases. We here focus on the dilute regime where particle interactions are  \textit{weak}. This implies that the gas either behaves as a non-interacting system or as an interacting  system on the ``ergodic" side of the many-body localization transition, where interactions typically lead to thermalization at long time.  In Sec. \ref{Sec:WL}, we first recall how weak localization naturally emerges in non-equilibrium momentum distributions of an ultracold, non-interacting gas. In Sec. \ref{Sec:CFS}, we extend this discussion to the regime of Anderson localization, which, in momentum space, drives another interference mechanism known as coherent forward scattering. Sec. \ref{Sec:position_space} then explains how these phenomena manifest themselves in the dual non-equilibrium configuration of a wave packet spreading in position space. In Sec. \ref{Sec:interactions}, finally, we discuss how interactions affect the out-of-equilibrium dynamics of a disordered Bose gas at the mean-field level, by playing the role of a dephasing mechanism for weak localization effects, promoting the emergence of thermalization or driving the gas to a pre-thermal superfluid regime.

\section{Weak localization in non-interacting quantum gases}
\label{Sec:WL}

A practical strategy for measuring weak localization of cold atoms was initially proposed in \cite{Cherroret12}, directly inspired of the experiments carried out in optics where weak localization is detected through the coherent backscattering (CBS) effect \cite{Albada85, Wolf88}. In the optical context, CBS is usually observed by sending a beam of well-defined direction $\bk_0$ into a disordered material and looking at the angular distribution of the reflected signal. In the language of cold atoms, having a well-defined direction means preparing an atomic cloud in a state $|\Psi(t=0)\rangle\simeq|\bk_0\rangle$, which can be done by communicating a mean momentum to the gas while making its dispersion of momenta around $\bk_0$ very small. Information about the angular distribution of scattered particles is then contained in the disorder-averaged momentum distribution $\smash{n_\bk(t)\equiv\overline{|\langle\bk|\Psi(t)\rangle|^2}}$. 
This suggests the following experimental protocol. A gas is prepared in the state $|\bk_0\rangle$, let evolve in a spatially disordered potential and finally imaged by time of flight at the desired time. 
\begin{figure}[h]
\centering\includegraphics[scale=0.14]{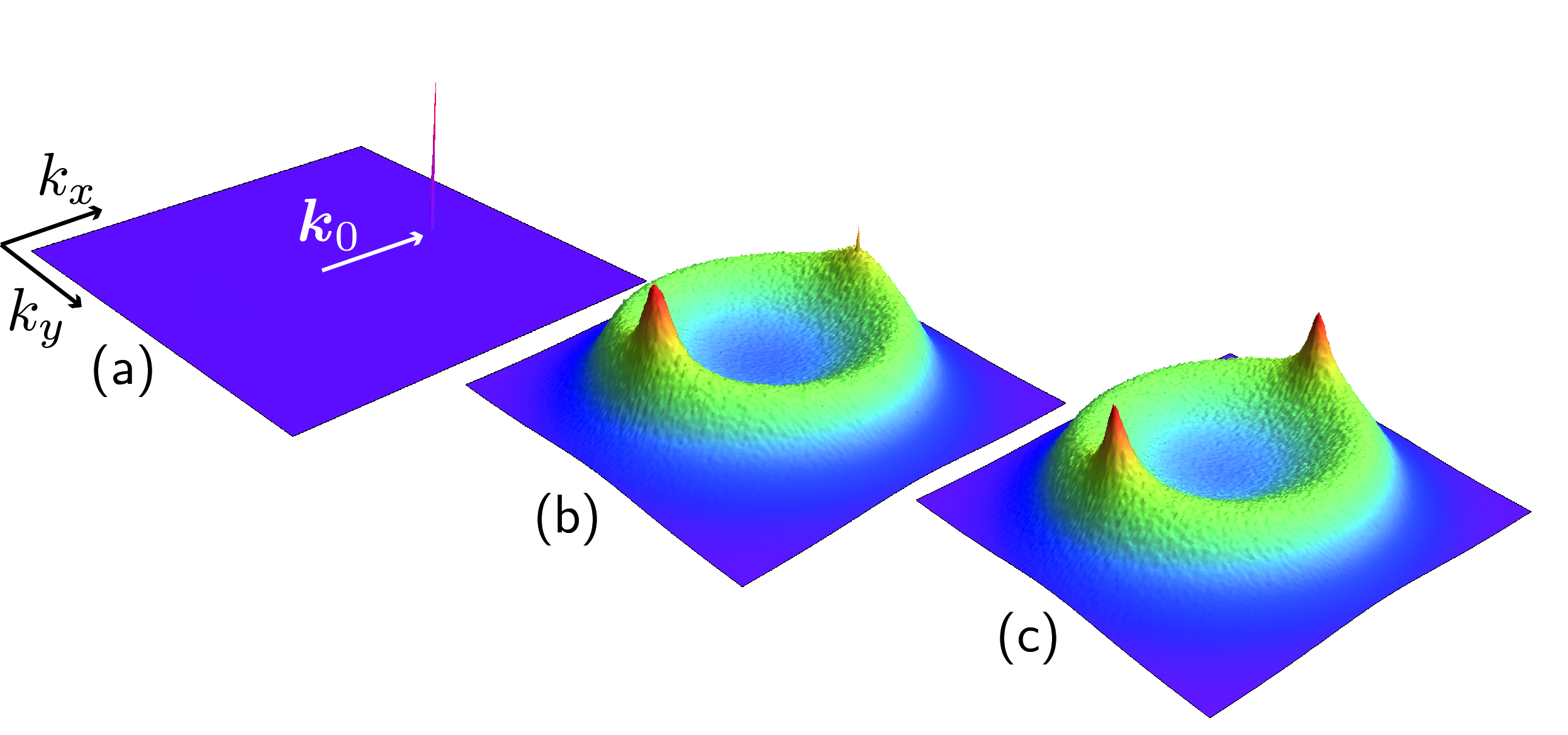}
\caption{\label{CBS_CFS_fig} 
Average momentum distribution $n_\bk(t)$ obtained after numerical propagation of a plane wave $|\bk_0\rangle$ with the time-dependent Schr\"odinger equation in a 2D random potential $V(\br)$ with speckle statistics and correlation function $\overline{V(\br)V(\br')}=[2V_0J_1(|\br-\br'|/\sigma)/(|\br-\br'|/\sigma)]^2$ ($\sigma$ is the speckle correlation length). Chosen parameters are $k_0=1.5/\sigma$ and $V_0=5/(m\sigma^2)$, and the propagation scheme involves an additional energy filtering aimed at selecting a well defined energy (here $\epsilon=-2.5$) for localized atoms \cite{Ghosh14}. The plots show three different times: (a) $t=10m\sigma^2$, only the initial mode is visible  (b) $t=30m\sigma^2$, diffusive regime, a CBS peak has emerged on top of the diffusive ring, and (c) $t=3500 m\sigma^2$, Anderson localization regime, both CBS and CFS peaks are present and symmetric. Data are averaged over some 7200 disorder realizations.
}
\end{figure}
Fig. \ref{CBS_CFS_fig} shows the distribution $n_{\bk}(t)$ obtained by numerically evolving in time
a plane-wave state $|\bk_0\rangle$ with the Hamiltonian $H=\boldsymbol{p}^2/2m+V(\br)$, choosing a speckle statistics for the random potential $V(\br)$. The distributions correspond to three different times.
At $t=0$, $n_\bk$ consists of a delta peak centered on $\bk_0$. As particles are scattered on the random potential, this initial mode gets quickly  depleted as $\exp(-t/\tau)$. Scattering occurs at a rate governed by the mean free time $\tau\equiv\ell/(k_0/m)$, where $\ell$ is the mean free time, i.e. the typical distance between two consecutive scattering processes (here and in the following, we set $\hbar=1$). 
These scattered atoms populate all other accessible $k$-space modes and, after a few  $\tau$, lose the memory and their initial direction \cite{Plisson13} and distribute within a time-independent ring $n^D_\bk$ of radius $|\bk| = k_0$ (Fig. \ref{CBS_CFS_fig}(b)). This ring  has a finite width $\sim 1/\ell$ due to the  dispersion of random potential's energies. Precisely \cite{Cherroret12}:
\begin{align}
\label{eq:ndk}
n^D_\bk=\int_{-\infty}^\infty {\rm d}\epsilon\, \frac{A_\epsilon(\bk)A_\epsilon({\bk_0})}{\nu_{\epsilon}}\simeq \frac{1}{\pi\nu_{\epsilon_0}\tau}\frac{1}{(\epsilon_0-\epsilon_k)^2+1/\tau^2},
\end{align}
where $A_\epsilon(\bk)\equiv \overline{\langle\bk|\delta(E-H)|\bk\rangle}$ is the spectral function of a particle in the random potential and $\nu_\epsilon$ is the density-of-states (dos) per unit volume. 
As long as $t$ does not exceed the localization time (see Sec. \ref{Sec:CFS}), multiple scattering essentially reduces to a \emph{diffusive} process. In this regime, the spectral function is given by $A_\epsilon(\bk)=1/(2\pi\tau)\times[(\epsilon-\epsilon_k)^2+1/(4\tau^2)]^{-1}$, where $\epsilon_k=\bk^2/2m$ and $\epsilon_0=\bk_0^2/2m$, yielding the second equality in Eq. (\ref{eq:ndk}).
The interpretation of Eq. (\ref{eq:ndk}) is the following. After the quench, at $t=0^+$, the quantum gas prepared in the state $|\bk_0\rangle$ acquires an energy distribution $A_\epsilon(\bk_0)$ in the random potential. Since scattering is elastic, this distribution remains constant during the time evolution. When the measurement is performed, finally, a particle with energy $\epsilon$ is detected with momentum $\bk$ with the probability density $A_\epsilon(\bk)/\nu_\epsilon$. 

Fig. \ref{CBS_CFS_fig}(b) also shows that a CBS peak emerges in the distribution around $\bk=-\bk_0$. The CBS peak is associated to the weak-localization interference between two time-reversed multiple scattering paths (see Fig. \ref{diagrams_CFS}(b) below). At exact backscattering, this interference is fully constructive and leads to a doubling of the momentum distribution:
\begin{align}
\label{eq:ndck}
n_{-\bk_0}=2n^D_{-\bk_0}.
\end{align}
 In practice,  detecting a well contrasted CBS peak requires the dispersion $\Delta k$ of particle momenta to be much smaller than the CBS width, which is of the order of $1/\ell$. Typical disordered optical potentials used in cold-atom setups 
have mean free paths of the order of a few $\mu$m \cite{Fallani08}. 
For a thermal gas of Rubidium atoms ($\Delta k\sim \sqrt{2m k_\text{B}T}$), the condition $\Delta k\ll \ell^{-1}$ leads to a temperature $T\lesssim$ few nK. Therefore, a clean observation of CBS requires to use an \emph{ultracold} cloud. This task was accomplished experimentally in \cite{Jendrzejewski12}, starting from a non-interacting Rubidium Bose-Einstein condensate. The gas was initially subjected to a brief harmonic pulse so to narrow its momentum distribution down to $\Delta k=0.12 $ mm.s$^{-1}$. Another observation of atomic CBS was also reported \cite{Labeyrie12}. In this experiment, however, no harmonic pulse was initially applied to the atomic coud, so that significant position-momentum correlations were present in the gas. In \cite{Cherroret13}, these correlations were shown to be responsible for another, purely classical ``backscattering echo'' showing up on top of the CBS signal.

\section{Coherent forward scattering effect}
\label{Sec:CFS}

\subsection{Diffusive regime}

A close look at the distribution in Fig. \ref{CBS_CFS_fig}(b) reveals that, at intermediate time, the momentum distribution is not only characterized by a diffusive ring and a CBS peak, but also exhibits a  smooth bump around $\bk=+\bk_0$. This bump is called the coherent forward scattering (CFS) peak. It was first discovered and characterized in \cite{Karpiuk12, Ghosh14} for matter waves, though signatures of enhanced forward scattering  were also noticed in the context of excitons in semiconductor nanostructures \cite{Runge02}. To understand its origin, let us recall a few qualitative elements of the semi-classical description 
of wave scattering in disorder. 
\begin{figure}[h]
\centering\includegraphics[scale=0.45]{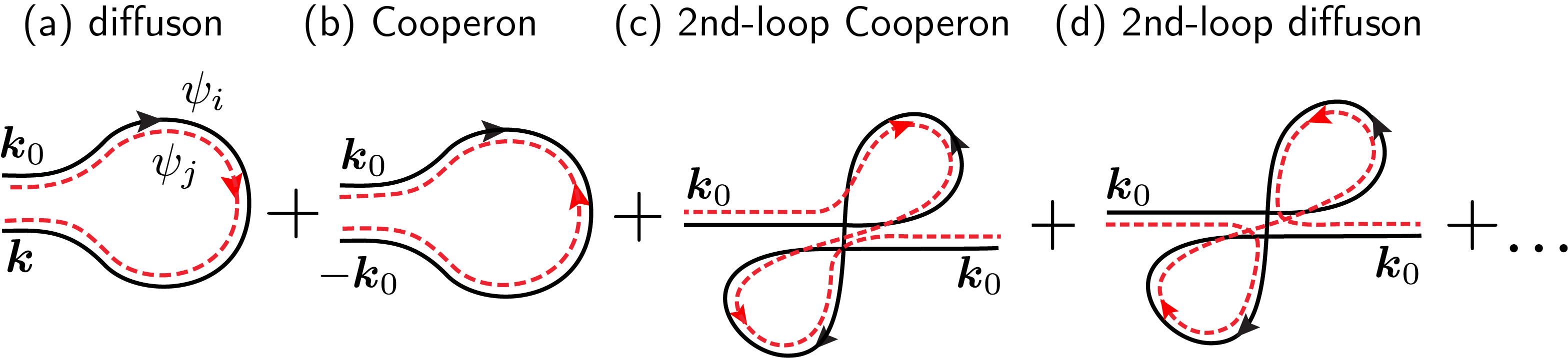}
\caption{\label{diagrams_CFS} 
Diagrams describing (a) classical diffusion  and  (b) coherent backscattering. (c) and (d) are the leading-order diagrams contributing to the coherent forward scattering peak.}
\end{figure}
In the diffusive regime, the disorder-averaged momentum distribution $n_\bk(t)\sim \sum_{i,j}\overline{\psi_i\psi_j^*}$ can be expressed as a weak-localization perturbation expansion involving  interference between two multiple scattering paths $\psi_i$ and $\psi_j$. The zeroth-order term of this expansion describes a co-propagation of the paths along the same sequence of scattering events, schematized by the diagrams of Fig. \ref{diagrams_CFS}(a). This contribution does not involve any interference and leads to the diffusive ring in Fig. \ref{CBS_CFS_fig}(a) and to Eq. (\ref{eq:ndk}). The leading-order correction is the first-order weak localization diagram shown in Figs. \ref{diagrams_CFS}(b), in which the paths $\psi_i$ and $\psi_j$ respectively propagate along the same scattering sequence  but in opposite directions. It is responsible for the CBS peak in Fig. \ref{CBS_CFS_fig}(b). The contributions (a) and (b) are known as ``diffuson'' and ``Cooperon''. 
At next order, the two new types of interference processes shown  in Figs. \ref{diagrams_CFS}(c) and (d) appear \cite{Hikami81}. These processes are qualitatively different in that they yield a contribution peaked around $\bk=+\bk_0$. Indeed, diagram (c) involves the direct concatenation of two Cooperons, such that an atom with initial momentum $\bk_0$ is twice  backscattered and ends up with a momentum $\bk_0$ after the multiple scattering sequence. The other correction (d) is the time-reversed version of diagram (c): both are thus  strictly equal in a time-reversal invariant system, but only diagram (d) survives when this symmetry is broken. 

The two-loop interference diagrams (c) and (d) constitute the leading-order contributions to the CFS peak. This was initially postulated in \cite{Karpiuk12} and later proven rigorously in \cite{Micklitz14} on the basis of the nonlinear $\sigma$-model in a quasi-1D geometry. The two-loop structure of the CFS interference can be used to qualitatively estimate the magnitude of the CFS peak at short time. To this end, let us invoke a simple geometrical argument where a scattering trajectory in the random potential is seen as a semi-classical ``tube'' of cross-section $\lambda_0^{d-1}$ and length $v_0t$ ($v_0=k_0/m$, $\lambda_0=2\pi/k_0$). The probability for the processes in Figs. \ref{diagrams_CFS}(c)  and (d) to occur is then essentially the geometrical probability for the Cooperon loop (b) to \textit{cross itself} at some point along the trajectory. This probability $P_\text{2-loop}\sim (\lambda_0^{d-1} v_0t)/V_\text{tot}$ is given by the ratio of the tube volume to the total volume accessible to the particles. At time $t\gg\tau$ and in the diffusion regime, the latter is $V_\text{tot}\sim(Dt)^{d/2}$, so that:
\begin{equation}
\label{P2loop_eq}
P_\text{2-loop}\sim \dfrac{1}{(k_0\ell)^{d-1}}\left(\dfrac{t}{\tau}\right)^{1-d/2}.
\end{equation}
This prediction reproduces the result of a rigorous, microscopic calculation \cite{Karpiuk12}, and shows that in two and three dimensions the CFS peak contrast is always small in the diffusive regime, as is indeed observed numerically in Fig. \ref{CBS_CFS_fig}(b).

\subsection{Anderson localization regime}
\label{Sec:CFS_loc}

The fate of the CFS peak in the Anderson localization regime can be assessed by extrapolating the semi-classical formula $P_\text{2-loop}\sim (\lambda_0^{d-1} v_0t)/V_\text{tot}$. In the  localization regime, the volume accessible to the particles becomes restricted to the localization length $\xi$, so that $V_\text{tot}\sim\xi^d$ and
\begin{equation}
\label{P2loop_eq_loc}
P_\text{2-loop}\sim \dfrac{k_0\ell}{(k_0\xi)^d}\dfrac{t}{\tau}.
\end{equation}
Although not rigorous, this equation suggests that the CFS peak grows in the localization regime, and that its contrast becomes of the order of unity at the time
\begin{align}
\label{tloc_def}
t\sim m k_0^{d-2}\xi^d\sim \nu\xi^d\equiv t_\text{loc},
\end{align}
which is nothing but the time scale on which Anderson localization shows up in dimension $d$. Such a growth of the CFS peak is well observed numerically, as seen in Fig. \ref{CBS_CFS_fig}(c) which shows the  momentum distribution well beyond the localization time. In practice, the CFS peak grows until it becomes symmetric to the CBS peak and the distribution becomes fully independent of time.

To understand the long-time structure of  Fig. \ref{CBS_CFS_fig}(c), the semi-classical perturbation theory --strictly valid when $t\ll t_\text{loc}$-- is no longer useful. 
Nonetheless, some insight about the limit $t\gg t_\text{loc}$ can be inferred from a simple non-perturbative argument based on the expansion of the state vector $|\Psi(t)\rangle$ of the gas over the basis of  localized eigenstates, $|\Psi(t)\rangle=\sum_n \langle\phi_n|\bk_0\rangle e^{-iE_nt}|\phi_n\rangle$ \cite{Lee14, Ghosh14}. 
From this expansion it follows that:
\begin{align}
\label{nbk_expansion}
n_\bk(t)\!=\!\sum_{m,n}
\overline{\phi_n^*(\bk_0)\phi_m(\bk_0)\phi_n(\bk)\phi_m^*(\bk)e^{-i(E_n-E_m)t}}
\!\underset{t\gg t_\text{loc}}{\simeq}\!
\sum_n\overline{|\phi_n(\bk_0)|^2|\phi_n(\bk)|^2}.
\end{align}
To write the second equality, we explicitly used that at long time particle motion becomes confined to a volume $\xi^d$  due to Anderson localization. In this regime, localized eigenstates are typically separated by $|E_n-E_m|\sim (\nu\xi^d)^{-1}\equiv 2\pi/t_\text{loc}$. Therefore, when  $t\gg t_\text{loc}$ the off-diagonal phase factors $(E_n-E_m)t$ oscillate very fast and va\-nish after disorder averaging. 
The long-time structure of $n_\bk(t)$ follows from the simple result that the localized eigenstates $\phi_n(\bk)$ are complex random Gaussian variables\footnote{This statistical property can be inferred from the spatial structure of localized modes: $\phi_n(\br)\sim e^{i\varphi_n(\br)}e^{-|\br|/\xi}$, where the random phases $\varphi_n(\br)$ fluctuate at the scale of $\ell\ll\xi$.}, uncorrelated at different $\bk$. Furthermore, due to  time-reversal invariance the additional symmetry relation $\phi_n(\bk)=\phi_n^*(-\bk)$ holds. Put together, these two properties lead to:
\begin{align}
\label{nbk_inf_CFS}
n_{\bk_0}(t\gg t_\text{loc})=
n_{-\bk_0}(t\gg t_\text{loc})=
2n_{\bk\ne \pm\bk_0}\equiv 2n^D_\bk.
\end{align}
In other words, the asymptotic distribution consists of a smooth background $n^D_\bk$ and of two peaks at $\bk=\pm\bk_0$, in agreement with Fig. \ref{CBS_CFS_fig}.
In passing, time-reversal invariance also implies that $n_{\bk}(t\gg t_\text{loc})=n_{-\bk}(t\gg t_\text{loc})$ for all $\bk$, which confirms the symmetric structure of the distribution at long time.

\begin{figure}
\centering\includegraphics[scale=0.5]{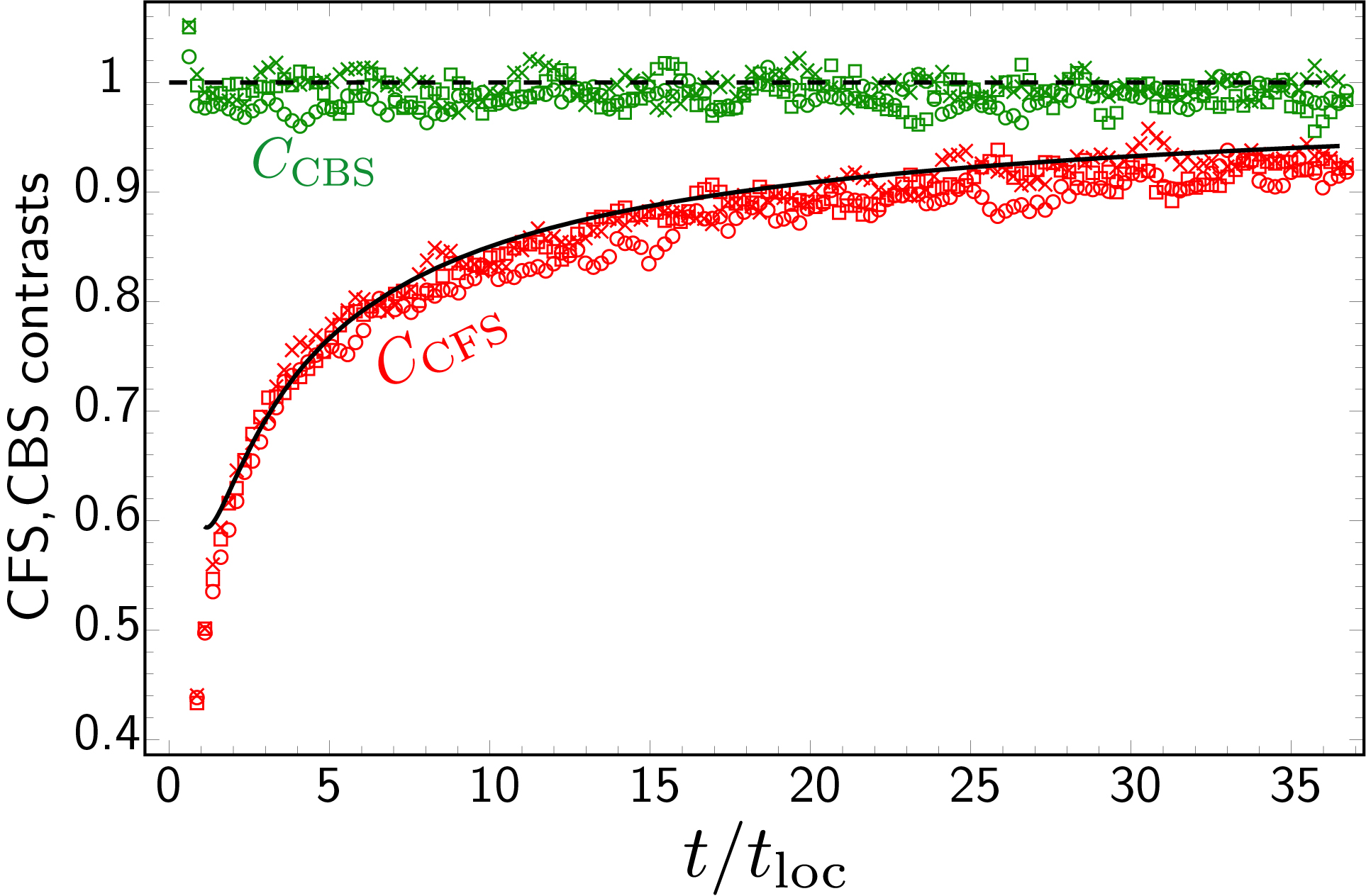}
\caption{\label{CBSCFS_strong}
Contrast of the CBS (green symbols) and CFS (red symbols) peaks as a function of $t/t_\text{loc}$, obtained from numerical simulations in a 2D speckle potential \cite{Ghosh14} (parameters are the same as in Fig. \ref{diagrams_CFS}). The solid curve is a fit to Eq. (\ref{CFS_contrast_theo}), using $\alpha$ and $t_\text{loc}$ are as fit pa\-ra\-me\-ters. The dashed line is the theoretical prediction $C_\text{CBS}=1$ corresponding to Eq. (\ref{nbk_inf_CFS}).
}
\end{figure}

We now have an almost complete picture of the time evolution of the CFS  peak from the diffusive to the localized regime: its contrast, $C_\text{CFS}(t)\equiv (n_{\bk_0}(t)-n_{\bk_0}^D)/n_{\bk_0}^D$, slowly grows according to Eq. (\ref{P2loop_eq}) when $t\ll t_\text{loc}$, and eventually saturates at 1 when $t\gg t_\text{loc}$, Eq. (\ref{nbk_inf_CFS}). 
The question of its precise time dependence in between these two limits is more tricky. It can be accessed by inserting the dos fluctuations $\delta\rho=\rho-\overline{\rho}$ in the mode expansion (\ref{nbk_expansion}), assuming a statistical independence of eigenfunctions and eigenenergies \cite{Lee14, Ghosh14}. This procedure leads to
\begin{align}
\label{nbk_CFS_K}
C_\text{CFS}(t)=
\int d\omega\, e^{-i\omega t}
\frac{\overline{\delta\rho_{\epsilon_0}\delta\rho_{\epsilon_0+\omega}}}{\overline{\rho_{\epsilon_0}}}.
\end{align}
Eq. (\ref{nbk_CFS_K}) underlines an interesting connection between the CFS peak contrast and the Fourier transform of the correlator of dos fluctuations, a quantity known as the form factor. Calculation of this quantity is still a complicated theoretical problem in general. In quasi-1D systems, an analytical result was recently obtained on the basis of the nonlinear sigma-model  \cite{Micklitz14, Marinho18}. In dimensions two and three, no exact result is available but an approximate expression can be found when $t\gg t_\text{loc}$, based on a Mott argument. Indeed, in this limit the form factor is dominated by the correlation of nearby levels in the spectrum ($|\omega|\ll 1/\nu\xi^d$). This correlation can be estimated by diagonalizing the $2\times 2$ hybridization Hamiltonian coupling  such pairs of levels, with a coupling strength governed by the overlap of the two associated localized wave functions
\cite{Mott70, Sivan87, Altland14, Ghosh14}. In  dimension two, this approach gives:
\begin{align}
\label{CFS_contrast_theo}
C_\text{CFS}(t\gg t_\text{loc})\simeq1-\alpha\frac{\text{ln}(t/t_\text{loc})}{t/t_\text{loc}},
\end{align}
where $\alpha$ is a phenomenological prefactor whose precise determination would require a microscopic calculation. Fig. \ref{CBSCFS_strong} compares numerical results for the CFS contrast as a function of time with Eq. (\ref{CFS_contrast_theo}). 

Beyond 2D disorder, the CFS peak was also numerically studied in one dimension \cite{Lee14} and in 3D random potentials. In the latter case, its contrast was shown to undergo a jump in the vicinity of the Anderson transition, whose finite-time scaling gives access to the critical exponent of the Anderson transition \cite{Ghosh17}. At the mobility edge, the CFS peak contrast was also related to multifractal dimensions \cite{Ghosh17, Martinez20}.

\section{Coherent multiple scattering in position space: mesoscopic echo}
\label{Sec:position_space}

\subsection{Diffusion and localization of wave packets}
\label{sec_WP_spreading}

The ``dual'' version of the non-equilibrium configuration of the previous section corresponds to preparing a quantum gas in a narrow wave-packet state $|\br_0\rangle$ and following its evolution in position space. Localization phenomena in the random potential are then probed through the disorder-averaged density distribution $\smash{n_\br(t)\equiv\overline{|\langle\br|\Psi(t)\rangle|^2}}$. This scenario is frequently encountered in experiments. 
In the diffusive regime, the density distribution is given by the well-known Gaussian law
\begin{align}
n_\br(t)=n^D_\br(t)=\frac{e^{-(\br-\br_0)^2/2Dt}}{(4\pi D t)^{d/2}}.
\label{diff_eq_pedestal}
\end{align}
In the localization regime, on the other hand, the wave packet becomes time independent and its density is exponentially localized at large distance, 
\begin{align}
\label{loc_eq_pedestal}
n_\br(t)\sim e^{-|\br-\br_0|/\xi}.
\end{align}
In one dimension, localization is the rule and the diffusive regime (\ref{diff_eq_pedestal}) never occurs, the density qualitatively obeying Eq. (\ref{loc_eq_pedestal}) after a few mean free times. The temporal freezing of wave packets was the criterion exploited in the first experimental demonstrations of localization of cold atoms \cite{Billy08, Roati08}. 
In two dimensions, Eq. (\ref{diff_eq_pedestal}) holds at times $\tau\ll t\ll t_\text{loc}$, while Eq. (\ref{loc_eq_pedestal}) applies at long times $t\gg t_\text{loc}$, with the localization time given by Eq. (\ref{tloc_def}). In three dimensions, finally, the two regimes are separated by a critical energy, the mobility edge of the Anderson transition, where the atoms move sub-diffusively. At the onset of the transition, the disorder is typically very strong. In cold-atom experiments searching for 3D Anderson localization  \cite{Jendrzejewski12b, Semeghini15}, this implies that the quantum gases  have rather broad energy distributions, and thus consist of a complicated mixture of diffusive, localized and critical atoms. Localized atoms are then detected after the diffusive atoms have moved away. The broad energy distribution of the cloud nevertherless makes the transition difficult to identify \cite{Muller16}, which partly explains the remaining discrepancies between experiment and theory on the position of the mobility edge \cite{Pasek17}.

\subsection{Mesoscopic echo}
\label{meso_echo_theory}

\begin{figure}[h]
\centering\includegraphics[scale=0.4]{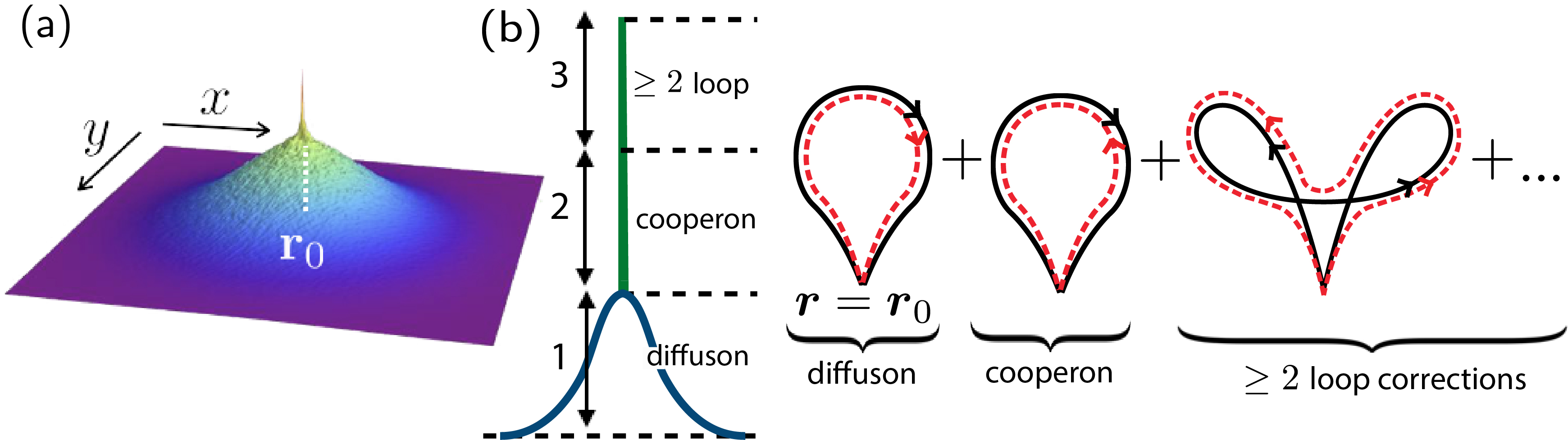}
\caption{
\label{ERO_2D_fig}
(a) Spatial distribution $n_\br(t)$ obtained by numerical propagation of a narrow wave packet in a 2D speckle potential after a few $\tau$. The mesoscopic echo is visible.
(b) Illustration of the profile of the density distribution at long time. When $t\ll t_\text{loc}$ the mesoscopic echo is governed by the Cooperon process, yielding an enhancement factor of 2. When $t\gg t_\text{loc}$, higher-loop diagrams proliferate, ultimately yielding an enhancement factor of 3, Eq. (\ref{nbr_inf_CFS}).
}
\end{figure}

In position space, exponential localization of wave packets is  not the only signature of coherent multiple scattering. Fig. \ref{ERO_2D_fig}(a) shows a typical profile $n_\br(t)$ obtained after a few mean free times
by numerically propagating a narrow wave packet in a 2D speckle potential with parameters close to those used for the plots in Fig. \ref{CBS_CFS_fig}.
The density distribution displays a broad pedestal and a narrow peak around the origin $\br=\br_0$. In the diffusive regime, the pedestal is well described by Eq. (\ref{diff_eq_pedestal}).
The peak at the origin, on the other hand, is a manifestation of weak localization in position space. This phenomenon, dubbed mesoscopic echo, was originally described in the context of quantum dots \cite{Prigodin94}. In the diffusive regime of a disordered system it occurs because the density probability for the particles to return to the origin point $\br_0$ is enhanced by the interference between time-reversed paths, as illustrated by the Cooperon diagram in Fig. \ref{ERO_2D_fig}(b). Due to time-reversal symmetry, this contribution equals its diffusive counterpart, the first diagram in Fig. \ref{ERO_2D_fig}(b), so that the density at $\br=\br_0$ is exactly doubled:\begin{align}
\label{eq:ndcr}
n_{\br_0}(t)=2n^D_{\br_0}(t).
\end{align}
The description of the mesoscopic echo  in terms of a Cooperon must be revisited at the onset of Anderson localization $t\sim t_\text{loc}$, where interference diagrams involving more than one loop start to proliferate. One of them is shown in Fig. \ref{ERO_2D_fig}(b): it is the exact counterpart of the CFS diagram (c) in Fig. \ref{diagrams_CFS}.  In position space, such a higher-loop process makes the enhancement factor of the mesoscopic echo grow beyond 2. The fate of the peak in the limit $t\gg t_\text{loc}$ can be inferred from a mode decomposition over localized states, similar to that used in Sec. \ref{Sec:CFS_loc}: $|\Psi(t)\rangle=\sum_n \langle\phi_n|\br_0\rangle e^{-iE_nt}|\phi_n\rangle$. Inserting this expansion into the spatial density $n_\br(t)=\overline{|\langle\br|\Psi(t)\rangle|^2}$ and, as for  Eq. (\ref{nbk_expansion}), neglecting the oscillating phase factors beyond the localization time, we obtain
\begin{align}
\label{nbr_expansion}
n_\br(t\gg t_\text{loc})\simeq
\sum_n\overline{|\phi_n(\br_0)|^2|\phi_n(\br)|^2}.
\end{align}
We then use that the localized wave functions are random Gaussian variables, uncorrelated at different positions. This is the same argument as in momentum space, except for one detail: the spatial eigenstates are here \emph{real} due the time-reversal invariance of the disordered system, so that the $\phi_n(\br)$ can be viewed as \emph{real} Gaussian random variables. This implies:
\begin{align}
\label{nbr_inf_CFS}
n_{\br_0}(t\gg t_\text{loc})=
3n_{\br\ne \br_0}.
\end{align}
In other words, from $t\sim t_\text{loc}$ onward, the mesoscopic echo grows beyond the factor-2 enhancement expected from the sole account of the Cooperon, and reaches a factor-3 enhancement at long times. This phenomenon, which is the counterpart of the growth of the CFS peak in momentum space, is schematized in Fig. \ref{ERO_2D_fig}(b). The growth of the mesoscopic echo in time is governed by the same curves as in Fig. \ref{CBSCFS_strong}: the factor-2 enhancement due to the Cooperon shows up over a few scattering times $\tau$, while the establishment of the factor 3 takes place over $t_\text{loc}$.

\subsection{Experimental observation}

The fully constructive interference (\ref{eq:ndcr}) and  (\ref{nbr_inf_CFS}) hold for an initially narrow wave-packet of well-defined de Broglie wavelength $\lambda_0=2\pi/k_0$. Under these conditions, the  width of the mesoscopic echo is of the order of $\lambda_0$. In practice however, the spatial distributions of cold gases prepared in optical traps are in general broader than $\lambda_0$, which leads to a reduction of the visibility of the mesoscopic echo. Even worse, the energy distribution of cold gases in random potentials is often not peaked around a well-defined energy $\hbar^2 k_0^2/2m$ but is instead broad, which further reduces the peak contrast. For these reasons, to our knowledge the mesoscopic echo was so far not seen in any cold-atom experiment involving disordered potentials. 

\begin{figure}[h]
\centering\includegraphics[scale=0.35]{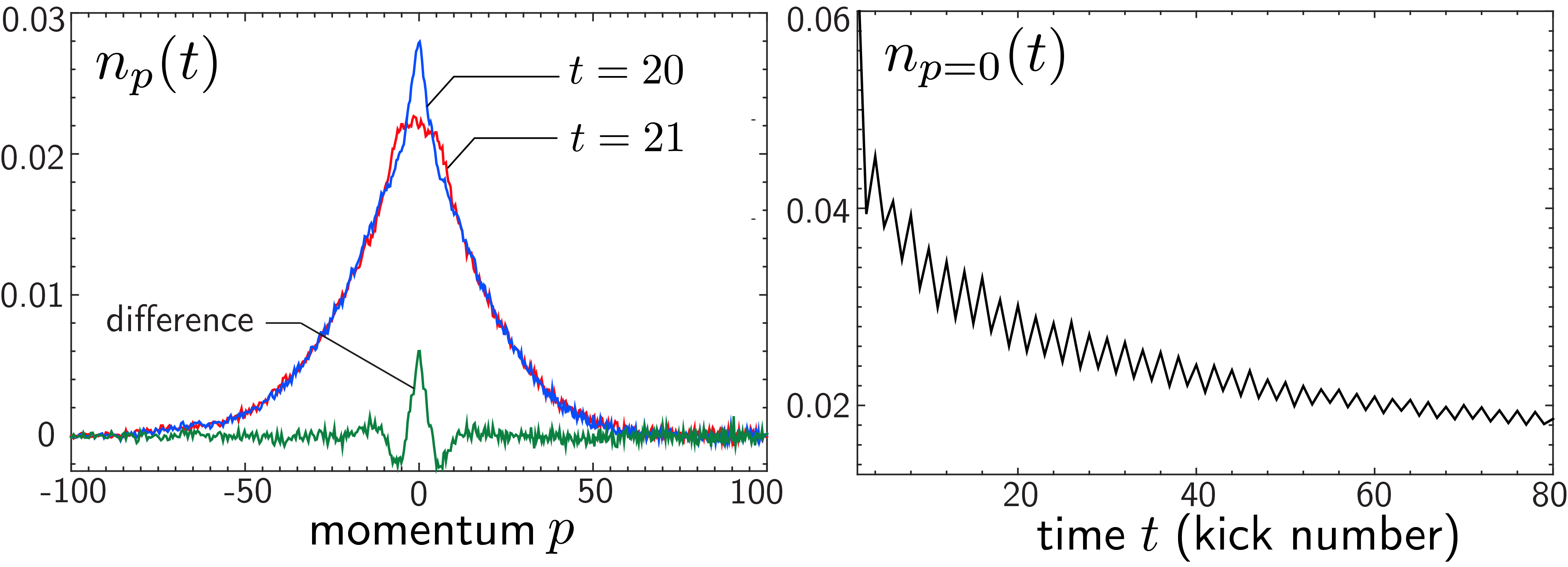}
\caption{
Left: experimental momentum distribution $n_p(t)$ measured in \cite{Hainaut17} at two successive kicks, using an atomic realization of the kicked rotor based on the Hamiltonian (\ref{eq_Hpsqkr}).
Right: Center of the wave packet $n_{p=0}(t)$ as a function of time, showing a periodic suppression and revival of the mesoscopic echo. 
\label{ERO_observation_fig}}
\end{figure}

The first successful observation of the mecoscopic echo in a quantum gas was achieved in an experimental realization of the atomic ``kicked rotor'' \cite{Hainaut17}. In this experiment, no spatial disorder was used, but rather a temporal form of disorder where the atoms are subjected to an optical standing wave modulated by a temporally-periodic sequence of  pulses. This setup, the atomic kicked rotor, has played an important role in the characterization of the  Anderson transition with cold atoms \cite{Chabe08, Lemarie10, Lopez12}. Its physics is the following. When subjected to a kick, an atom undergoes a change in momentum, whose value becomes quickly unpredictable after a few kicks. This corresponds to a chaotic motion, similar to a multiple scattering process but in momentum space. Like in spatially disordered systems, interference between scattering paths may also occur and lead to localization. In the atomic kicked rotor, the mesoscopic echo shows up around the point $p=0$ of the momentum distribution $\smash{n_p(t)\equiv \overline{|\langle p|\Psi(t)\rangle|^2}}$. Observing a well-contrasted peak in this system requires $n_p(t=0)$ to be as narrow as possible, i.e. a gas as cold as possible.
The experiment \cite{Hainaut17} used a gas a of Cesium atoms cooled down to $\sim 2\mu$K. Even at this low temperature though, the contrast of the mesoscopic echo is significantly decreased. For this reason, a \textit{differential} measurement was carried out, based on the following Hamiltonian
\begin{equation}
H\!=\!\frac{p^{2}}{2}+K\sum_{n}\left[\cos x\ \delta(t\!-\!2n)+\cos(x\!+\!a)\ \delta(t\!-\!2n\!+\!1)\right].
\label{eq_Hpsqkr}
\end{equation}
Eq.~(\ref{eq_Hpsqkr}) differs from the standard kicked-rotor Hamiltonian \cite{Casati79} owing to the additional spatial phase $a$ (the usual kicked rotor corresponds to $a=0$).
In  \cite{Hainaut17}, the blinking standing wave was produced by two counter-propagating lasers periodically switched on and off, the phase $a$ stemming from a spatial shift of the standing wave every second pulse. 
Experimental momentum distributions at two successive kicks are shown in the left panel of Fig. \ref{ERO_observation_fig}. The value of the distribution at $p=0$ is also shown in the right panel as a function of time.
Due to the phase shift $a$, the mesoscopic echo periodically disappears and reappears every second kick. This blinking is due to an additional phase $\exp(i\Phi)=\exp(i a\delta p )$ imprinted to an atom when its momentum is scattered by $\delta p$. For an even number of kicks, the direct and reversed paths accumulate the same total phase $\Phi_\text{dir}=\Phi_\text{rev}$, so that the interference is constructive, $\Phi\equiv \Phi_\text{dir}-\Phi_\text{rev}=0$. In contrast, for an odd number of kicks, the accumulated phases $\Phi_\text{dir}$ and $\Phi_\text{rev}$ have opposite signs, so that a finite phase difference $\Phi$ subsists: the interference is not constructive and the average over all paths suppresses the mesoscopic echo in general.
\begin{figure}[h]
\centering\includegraphics[scale=0.7]{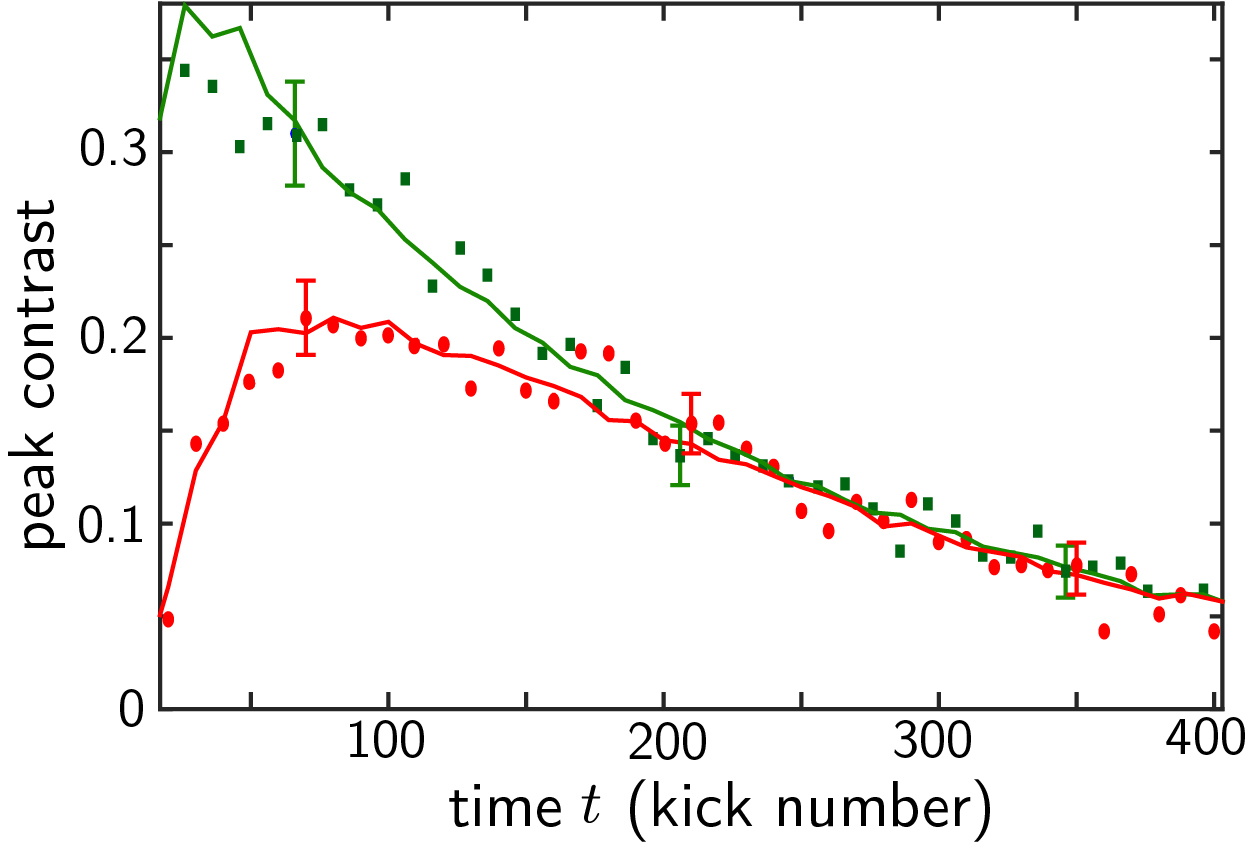}
\caption{
Contributions of the Cooperon (green) and of higher-loop diagrams (red) to the contrast of the mesoscopic echo, separately measured in the experiment \cite{Hainaut18}.
\label{ERO_loc_exp}}
\end{figure}

As explained in Sec. \ref{meso_echo_theory}, the enhancement factor of  the mesoscopic echo crosses-over from a factor 2 to a factor 3 at the onset of Anderson localization. At first sight, distinguishing experimentally between these two regimes seems hard since the visibility of the peak is anyhow strongly suppressed by the too broad size of the initial wave packet. A way to circumvent this difficulty was proposed in \cite{Hainaut18}, based on the idea that the Cooperon disappears when the temporal symmetry of the kick sequence is broken. Precisely, in \cite{Hainaut18} the kick amplitude $K$ in Eq. (\ref{eq_Hpsqkr}) was further periodically modulated in time. This modulation was chosen so to break the time-reversal symmetry of the kick sequence except at specific kicks where the Cooperon becomes nonzero. With this strategy  it was possible to temporally separate the contributions of the Cooperon and of higher-loop diagrams, and to access their individual time dependence. Those are shown in Fig. \ref{ERO_loc_exp}, which should be compared with the numerical prediction of Fig. \ref{CBSCFS_strong}. The difference with Fig. \ref{CBSCFS_strong} is explained by the presence of stray decoherence, which scrambles the mesoscopic echo exponentially at long times.


\section{Weak interactions in out-of-equilibrium disordered gases}
\label{Sec:interactions}

\subsection{Wave-packet spreading}

The effect of weak interactions on the dynamics of spreading wave packets in disorder (Sec. \ref{sec_WP_spreading}) has been explored in a number of works, with special interest for the case of 1D disordered chains \cite{Pikovsky08, Kopidakis08}. 
From a theoretical point of view, this problem is most easily tackled with bosons, which  at low temperature and in the mean-field approximation obey the nonlinear Gross-Pitaevskii equation
\begin{equation}
i\partial_t\Psi(\br,t)= \left[-\frac{\boldsymbol{\nabla}^2}{2m} + V(\br)+gN|\Psi(\br,t)|^2\right]
\Psi(\br,t)
\label{eq:grosspitaevskii}
\end{equation}
for the mean field $\Psi(\br,t)$ in a disordered potential $V(\br)$, with $g$ the interaction strength and $N$ the number of atoms. Here the field is normalized according to $\smash{\int d\br |\Psi(\br,t)|^2=1}$.
Despite the relative simplicity of the Gross-Pitaevskii equation, the problem turned out to be non-trivial, even in one dimension, at both the numerical and theoretical levels. 
In \cite{Pikovsky08, Kopidakis08}, it was numerically  found that for a small nonlinearity and strong disorder, wave packets are no longer localized at long time, but instead spread sub-diffusively in one-dimensional lattices, $\int d\br |\Psi(\br,t)|\br^2\propto t^\alpha$ with $\alpha<1$. In other words, Anderson localization is destroyed, but classical diffusion is not fully recovered. The precise value of $\alpha$ was somewhat debated \cite{Pikovsky08, Kopidakis08, Flach09, Garcia-Mata09, Iomin09, Mulansky13}, for the sub-diffusion process establishes very slowly and the precise estimation of $\alpha$ requires to run simulations over enormously long times (in the recent numerical work \cite{Vakulchyk18} aimed to determine $\alpha$, the time scale probed would correspond to $2.10^9$s in a real experiment!) Numerics nevertheless seems to suggest $\alpha=0.3 - 0.4$ in one dimension. The problem was also tackled in a random potential in three dimensions in the vicinity of the Anderson transition, where it was found that weak interactions also lead to sub-diffusion on the localized side of the transition, while they leave the critical point and the diffusive side of the transition essentially unaffected \cite{Cherroret14, Cherroret16}. The question of wave-packet spreading is not fully clarified though, since other works also put forward that sub-diffusion could be non-algebraic and eventually become slower than any power law at arbitrarily long times \cite{Wang09, Basko11}. Experimentally, signatures of sub-diffusion of a weakly interacting Bose-Einstein condensate were found in \cite{Lucioni11}.

\subsection{Thermalization and dephasing}
\label{Sec_mesoscopic_regime}

A different, and somewhat simpler out-of-equilibrium scheme addresses the effect of weak bosonic interactions following a quench from a plane-wave state. In this configuration, the average density $\smash{\overline{|\Psi(\br,t)|^2}}$ remains uniform at all times. This problem was originally examined in 1D disordered chains, where it was shown that the nonlinear coupling of localized modes always leads to a thermalization of the system at the mean-field level \cite{Kottos11, Basko11}. We here discuss it in the  2D scenario of Sec. \ref{Sec:WL}, where the Bose gas is prepared in the plane-wave state $|\bk_0\rangle$ and the dynamics at weak disorder is built upon diffusive modes. 

\begin{figure}[h]
\centering\includegraphics[scale=0.36]{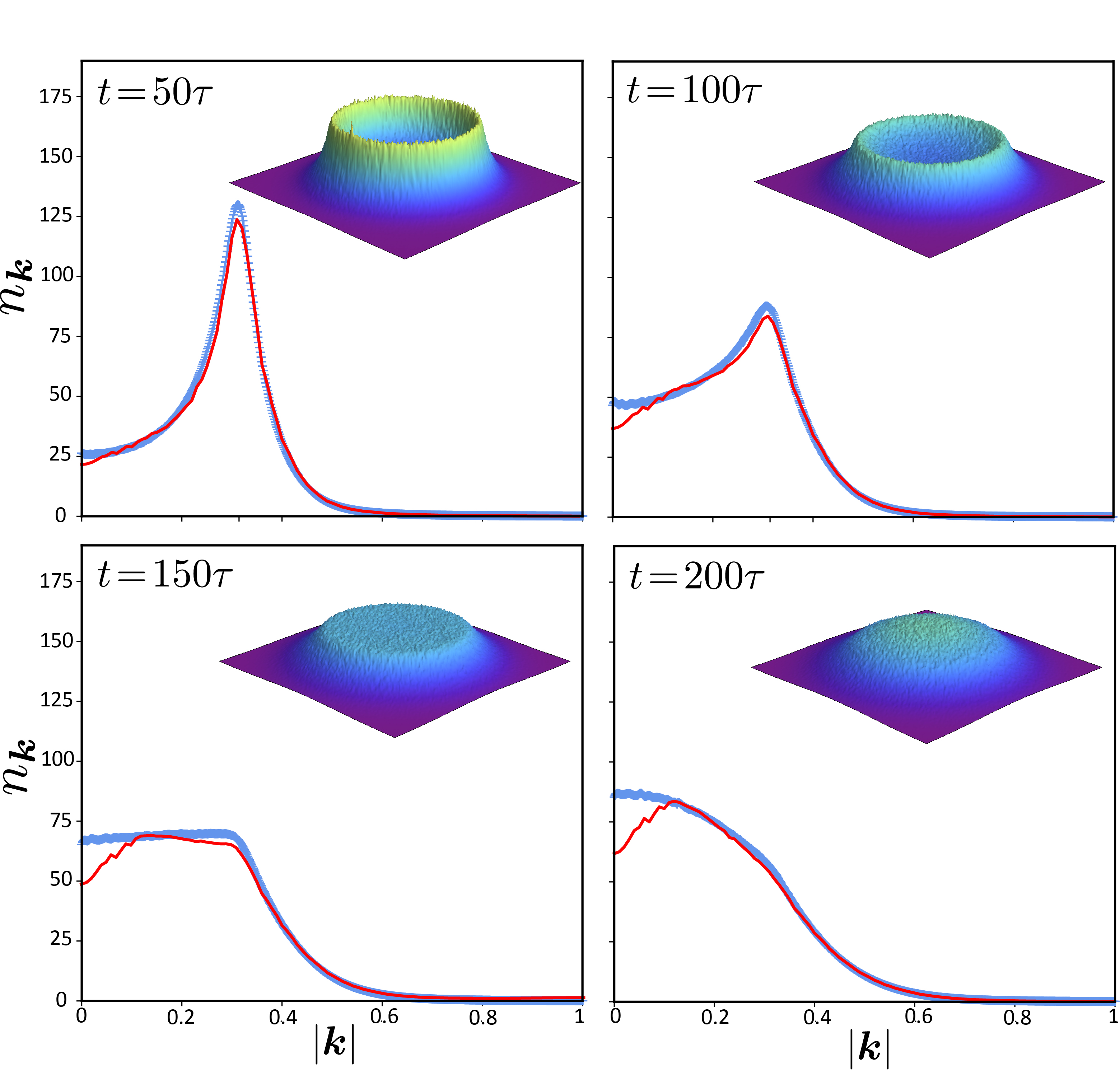}
\caption{
\label{momcut_vskinetic}
Radial cuts of the momentum distribution $n_\bk(t)$ of a Bose gas at four increasing times, obtained by numerically propagating the plane-wave state  $|\bk_0\rangle$ with the Gross-Pitaevskii equation (\ref{eq:grosspitaevskii}) in a Gaussian, uncorrelated random potential $V(\br)$ with $\overline{V(\br)V(\br')}=\gamma\delta(\br-\br')$. Insets show the corresponding density plots in the plane $(k_x,k_y)$. 
Chosen parameters are $k_0=0.314$ and $\gamma=0.01$, corresponding to a mean free time $\tau\simeq 100$, and the interaction energy  is set to $g\rho_0=0.002$. The  distributions involve an angular average and are further averaged over 1680 disorder realizations. 
Solid red curves are solutions  of the kinetic theory, Eqs. (\ref{eq:ndk_nonlinear}) and (\ref{eq:kinetic_eqD}), in which the maximum of the spectral functions is set to $\epsilon=\bk^2/2m$ (see main text).}
\end{figure}

When the initial state is a plane wave, the impact of interactions on the dynamics is entirely due to the random fluctuations of the nonlinear potential $gN|\Psi(\br,t)|^2$ in Eq. (\ref{eq:grosspitaevskii}). These fluctuations, which are time dependent, lead to inelastic particle collisions that modify the energy distribution of the Bose gas as it evolves in the disordered potential. At weak disorder these collisions occur with the time scale
\begin{align}
\label{nonlinear_time_def}
\frac{1}{\tau_\text{col}}\!\sim\!\int\!\frac{d^2\boldsymbol{k}}{2\pi}\overline{|\langle\boldsymbol{k}_0|gN|\Psi|^2|\boldsymbol{k}\rangle|^2}\delta(\epsilon_0-\epsilon_{\boldsymbol{k}})
\sim\frac{(g\rho_0)^2}{\epsilon_0}
\end{align}
where $\rho_0=N/V$ with  $V$ the volume of the system. 
In this configuration, the main effect of  particle collisions is to thermalize the momentum distribution $n_\bk(t)$. This can be seen in the insets of Fig. \ref{momcut_vskinetic}, which show numerical distributions obtained by solving the Gross-Pitaevskii equation (\ref{eq:grosspitaevskii}) in a random potential: the diffusive ring slowly evolves in time, and eventually becomes a smooth distribution centered on $\bk=0$.
A theoretical description of the thermalization process visible in the insets of Fig. \ref{momcut_vskinetic} was first developed in \cite{Schwiete13} based on a classical version of the Keldysh field theory, and later in \cite{Cherroret15, Scoquart20} by means on a non-pertubative diagrammatic technique built upon diffusion modes.
\begin{figure}[h]
\centering\includegraphics[scale=0.47]{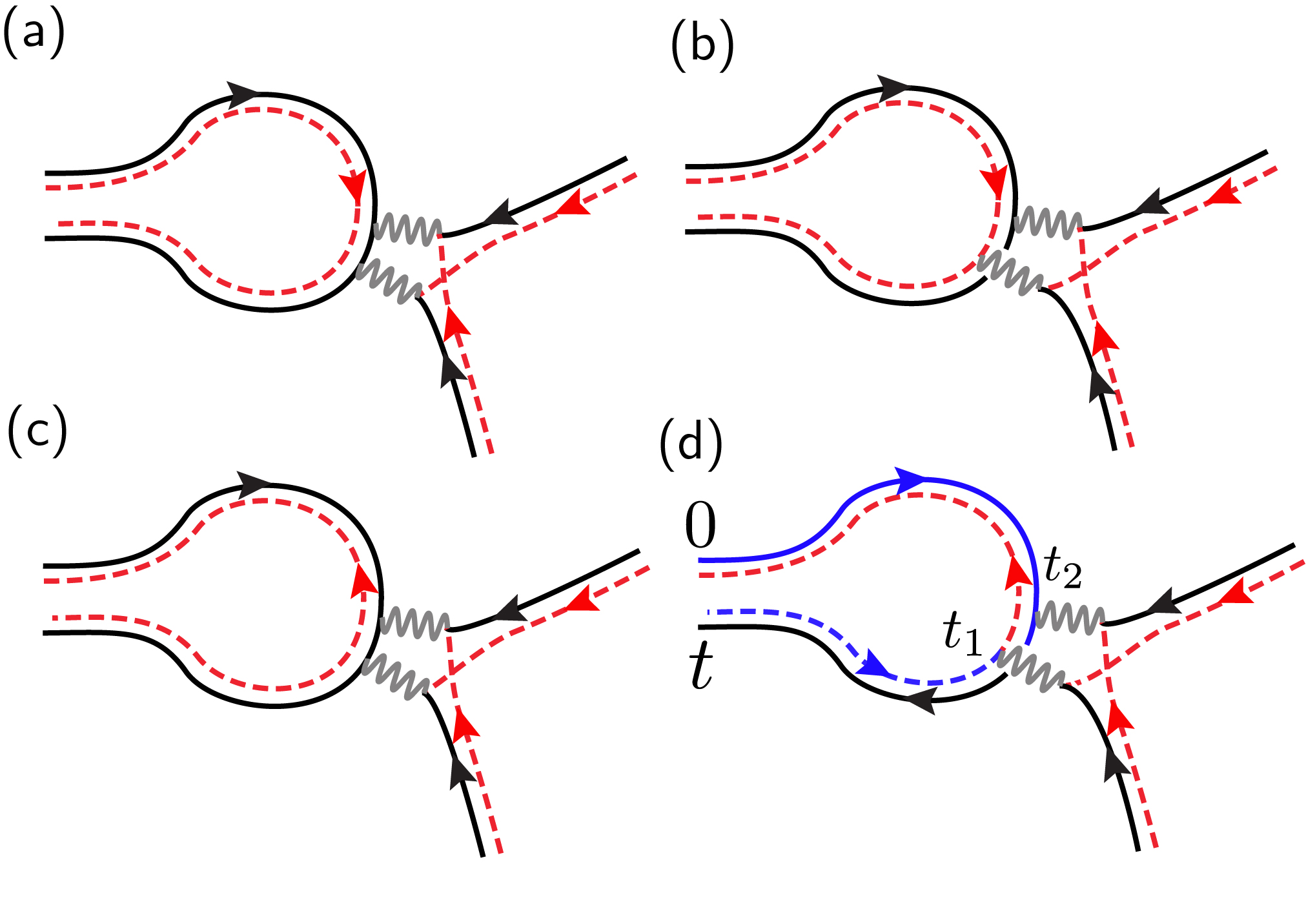}
\caption{
\label{nonlinear_diag.fig}
Typical elementary diagrams governing the dynamics of a weakly interacting, disordered Bose gas initially prepared in a plane-wave state $|\bk_0\rangle$. Upper diagrams are of diffuson type (two scattering paths, solid and dashed lines, co-propagate along an identical sequence of disorder scattering events) while lower ones are of Cooperon type (the paths propagate along the same scattering sequence but in opposite directions). In diagrams (a) and (c), a single sequence of the pair (here the solid one) is interrupted by two particle-collision events, represented by the wavy lines. On the contrary, in diagrams (b) and (d), both sequences of the pair are affected. In the case of diagram (d), this property  imposes the two collisions to occur at exactly half the sequence, i.e. at  $t_1=t_2=t/2$, which is very unlikely and makes this diagram negligible.}
\end{figure}
In the latter approach, it was shown that for $g\ne0$, the diffuson sequences of Fig. \ref{diagrams_CFS}(a) are modified by  nonlinear processes as those shown in Figs. \ref{nonlinear_diag.fig}(a) and (b). In such processes, a pair of multiple scattering paths is interrupted by two successive particle collisions occuring in between any two scattering events on the disordered potential. The collisions may affect a single path (diagram of type (a)) or both paths (type (b)) \cite{footnote}. A kinetic equation was then  derived from a diagrammatic resummation taking into account an arbitrary number of  such elementary processes along the diffusive sequence. As a net result, the diffusive component  (\ref{eq:ndk}) of the momentum distribution becomes, in the presence of interactions \cite{Scoquart20},
\begin{align}
\label{eq:ndk_nonlinear}
n^D_\bk=\int {\rm d}\epsilon\, A_\epsilon(\bk)f_\epsilon(t),
\end{align}
where $f_\epsilon(t)$ is, up to a density-of-state factor $\nu_\epsilon$, the energy distribution of the Bose gas, normalized according to $\int d\epsilon\, \nu_\epsilon f_\epsilon(t)=1$. In the absence of interactions, $f_\epsilon(t)=A_\epsilon(\bk_0)/\nu_\epsilon$ is independent of time and reduces to the spectral function of the gas in the disorder, so that Eq. (\ref{eq:ndk}) is recovered. When $g\ne0$ on the other hand, it becomes time dependent and obeys the Boltzmann-like kinetic equation
\begin{align}
&\partial_t f_\epsilon=\int_{\epsilon_1,\epsilon_2,\epsilon_3\geq0}\!\!\!\!\!\!\!\!\!\!\!\!\!\!\!{\rm d}\epsilon_1{\rm d}\epsilon_2~W(\epsilon,\epsilon_1,\epsilon_2)
\Bigl[(f_{\epsilon}+f_{\epsilon_1+\epsilon_2-\epsilon})f_{\epsilon_1}f_{\epsilon_2}-f_{\epsilon}f_{\epsilon_1+\epsilon_2-\epsilon}(f_{\epsilon_1}+f_{\epsilon_2})\Bigr]
\label{eq:kinetic_eqD}
\end{align}
with $\epsilon_3=\epsilon_1+\epsilon_2-\epsilon$ and the initial condition $f_\epsilon(0)=A_\epsilon(\bk_0)/\nu_\epsilon$.
The interaction kernel is given, in two dimensions and at weak disorder (Born approximation), by
\begin{align}
W(\epsilon,\epsilon_1,\epsilon_2)=\frac{m^3(g\rho_0)^2}{2\pi^4\nu_\epsilon}\frac{K\left(\frac{ 2\sqrt[4]{\epsilon\epsilon_1\epsilon_2\epsilon_3}}{\sqrt{\epsilon_1\epsilon_2}+\sqrt{\epsilon\epsilon_3}}\right)}{\sqrt{\epsilon_1\epsilon_2}+\sqrt{\epsilon\epsilon_3}},
\label{eq:kinetic_kernel}
\end{align}
where $K$ is the complete elliptic integral of the first kind. 
It should be noted that the integrals in Eqs. (\ref{eq:ndk_nonlinear}) and (\ref{eq:kinetic_eqD}) in principle run over all energies allowed by the dos. At the level of the Born approximation where the kernel $W$ is derived however, only positive energies are included \cite{Schwiete13, Scoquart20}, which leads to a poor estimation of  $n_\bk(t)$ at low values of $k$. To address this difficulty, an approximate method is to set the energy where the spectral functions $A_\epsilon(\bk)$ are maximum at $\epsilon=\bk^2/2m$ when computing Eq. (\ref{eq:ndk_nonlinear}) and the dos $\nu_\epsilon=\int d^2\bk/(2\pi)^2 A_\epsilon(\bk)$ in Eq. (\ref{eq:kinetic_kernel}). Roughly speaking, this approximation amounts to including a finite real self-energy into the theory (neglected at the Born approximation).
A comparison between the Gross-Pitaevskii simulations and Eqs. (\ref{eq:ndk_nonlinear}) and (\ref{eq:kinetic_eqD}) using this approach is shown in the main panels of Fig. \ref{momcut_vskinetic} at four increasing times. 
The agreement is very good, except at very low $k$-values.

The impact of interactions on the CBS peak in Fig. \ref{CBS_CFS_fig}(a) is slightly more tricky but, at weak disorder, it can also be tackled by means of a resummation of particle-collision diagrams. Typical scattering sequences to be considered are shown in Fig. \ref{nonlinear_diag.fig}(c) and (d). They are essentially the coherent counterparts of diagrams (a) and (b). The diagram (d), however, deserves a special attention. Let us assume that the particle collision on the direct path occurs at a time $t_1$ and the collision on the reversed path at $t_2$. As these two events are typically bound to occur within  a scattering mean free time, we have $t_1\simeq t_2$. Then, by definition of the time-reversed paths, $t_2$ also equals $t-t_1$ so that $t_1=t-t_1$. In other words, in such a process any particle-collision event is forced to occur exactly at half the sequence, $t_1=t/2$, which is extremely unlikely at long time. Diagrams of type (d) are therefore negligible, and only coherent diagrams of type (c) should be considered. 
Because of the ``missing'' diagrams of type (d),  
the contribution of the CBS peak  to the momentum distribution at $\bk=-\bk_0$ is not controlled by the distribution $f_\epsilon(t)$ but is instead given by
\begin{equation}
n^C_{-\bk_0}  =  \int{\rm d}\epsilon\, A_\epsilon(\bk_0)f^{C}_\epsilon(t),
\label{eq:mom_distrib_coop}
\end{equation}
where the new energy distribution  $f_\epsilon^C(t)\ne f_\epsilon(t)$ obeys the kinetic equation
\begin{align}
\partial_t f^C_\epsilon=f^{C}_{\epsilon}&\int_{\epsilon_1,\epsilon_2,\epsilon_3\geq0}\!\!\!\!\!\!\!\!\!\!\!\!\!\!\!{\rm d}\epsilon_1{\rm d}\epsilon_2~W(\epsilon,\epsilon_1,\epsilon_2)
\Bigl[f_{\epsilon_1}f_{\epsilon_2}-f_{\epsilon_1+\epsilon_2-\epsilon}(f_{\epsilon_1}+f_{\epsilon_2})\Bigr].
\label{eq:kinetic_eqC}
\end{align}
The asymmetry between the kinetic equations (\ref{eq:ndk_nonlinear}) and (\ref{eq:mom_distrib_coop}) implies that the CBS contrast is no longer one, but instead decays in time in the presence of interactions. Indeed, while Eq. (\ref{eq:ndk_nonlinear}) typically drives $f_\epsilon(t)$ to a time-independent Rayleigh-Jeans distribution at long time \cite{Cherroret15}, Eq. (\ref{eq:mom_distrib_coop}) on the contrary makes $f_\epsilon^C(t)$ evolve to zero. Inelastic particle collisions thus reduce to an effective \emph{dephasing} mechanism for weak localization. In spirit, this effect is similar to what is observed for thermal equilibrium electrons in good conductors, where electron-electron interactions lead to a finite coherence time that suppresses weak localization \cite{Altshuler65}. A difference is that the dephasing mechanism here occurs for an isolated system and in an out-of-equilibrium configuration. 
\begin{figure}[h]
\centering\includegraphics[scale=0.6]{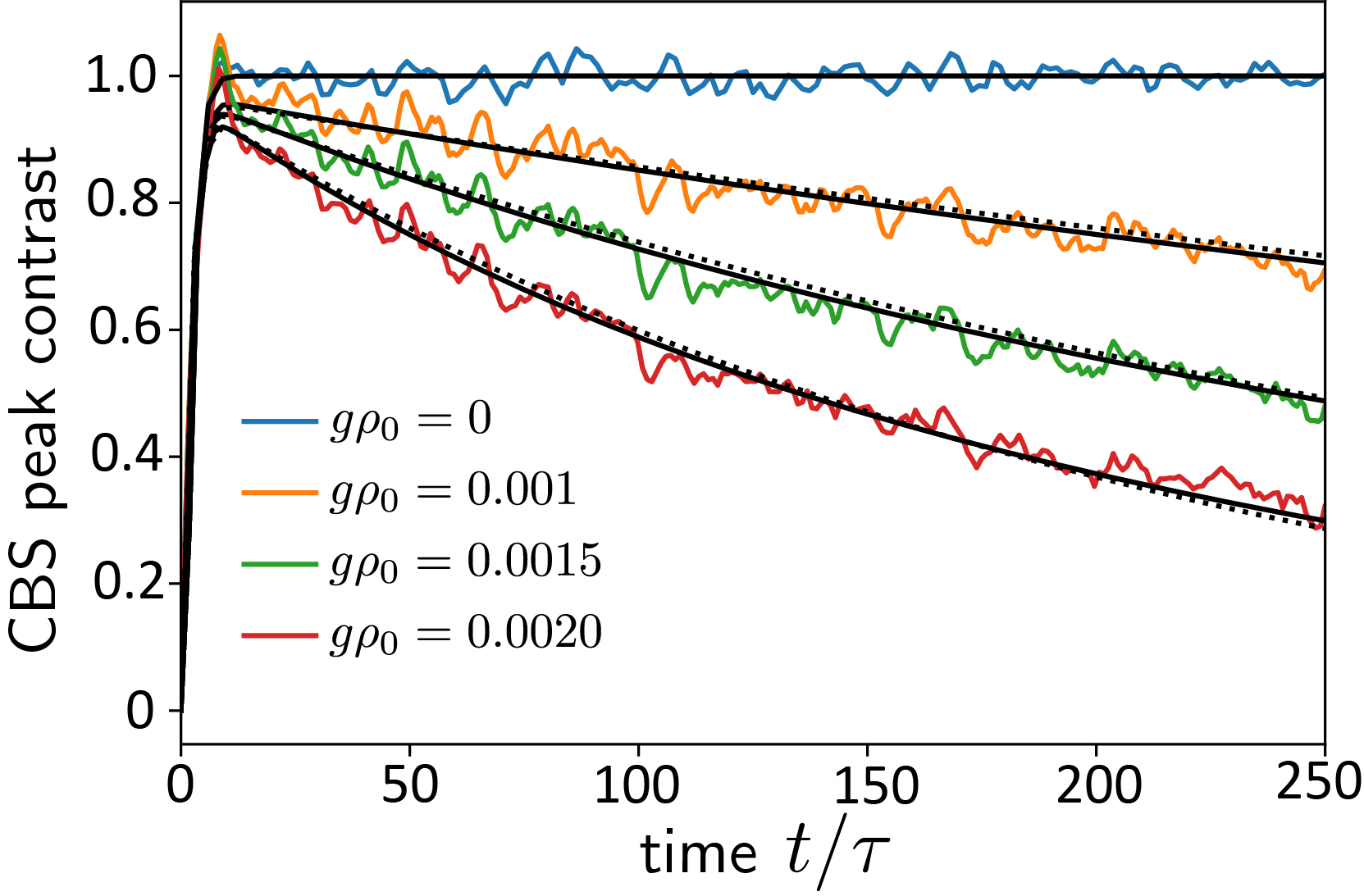}
\caption{
\label{CBS_contrast.fig}
Contrast of the CBS peak  versus time for increasing values of $g\rho_0$ from top to bottom (colored curves). As in Fig. \ref{momcut_vskinetic}, the data are obtained by propagating the plane wave $|\bk_0\rangle$ with the Gross-Pitaevskii equation (\ref{eq:grosspitaevskii}) using a Gaussian, uncorrelated random potential $V(\br)$ with $\overline{V(\br)V(\br')}=\gamma\delta(\br-\br')$, here for $k_0=0.628$
and $\gamma \simeq 0.0182$,  corresponding to a mean free time $\tau\simeq 51.84$. Data are averaged over about 16000 disorder realizations.
Solid smooth black curves are fits to the theory, Eqs. (\ref{eq:ndk_nonlinear}), (\ref{eq:mom_distrib_coop}), (\ref{eq:kinetic_eqD}) and (\ref{eq:kinetic_eqC}), using $g$ as a fit parameter \cite{Scoquart20}.
Dotted curves are the exponential law (\ref{Constrast_exponential_decay}).
}
\end{figure}
Fig. \ref{CBS_contrast.fig} shows the contrast of the CBS peak as a function of time, numerically computed from the Gross-Pitaevskii  equation, for increasing values of the interaction strength. The numerical data are well described by the kinetic approach described above, shown as solid curves. While there is no obvious analytical solutions of the kinetic equations (\ref{eq:kinetic_eqD}) and (\ref{eq:kinetic_eqC}) at finite time, we find the the decay of the CBS contrast is rather well captured by an exponential law:
\begin{align}
\label{Constrast_exponential_decay}
\frac{n^C_{-\bk_0(t)}}{n^D_{-\bk_0(t)}}
=\exp(-t/\tau_\phi)
\end{align}
where $\tau_\phi=5(g\rho_0)^2/\epsilon_0$ is the time scale governing the effective dephasing of CBS. Quite naturally, this time scale is found to be directly proportional to the particle collision time $\tau_\text{col}$, defined through Eq. (\ref{nonlinear_time_def}).

\subsection{Emergence of superfluidity}

In the previous section, we discussed an out-of-equilibrium regime where particles essentially experience multiple scattering, and where interactions are so weak that they thermalize the system and dephase weak localization effects over a time scale much longer than the scattering mean free time, $\tau_\text{col}\gg\tau$. 
In terms of microscopic energy scales, in such a ``mesoscopic'' regime one typically has $\epsilon_0=k_0^2/2m>V_0>g\rho_0$, where $V_0$ is the magnitude of disorder  fluctuations and $g\rho_0$ is the mean interaction energy. Since multiple scattering is the dominant mechanism, in this regime the Bose gas has a very poor spatial coherence. This coherence is encoded  in the first-order correlation
 \begin{equation}
 \label{g1_J0}
 g_1(\br,t)\equiv V\overline{\Psi^*(0,t)\Psi(\br,t)}=\!\int\!\!\frac{d^2\bk}{(2\pi)^2}n_\bk(t)e^{i\bk\cdot \br}.
 \end{equation}
When $g=0$, $n_\bk$ is approximately given by Eq. (\ref{eq:ndk}), which leads to
 \begin{equation}
 \label{g1_nointeractions}
 g_1(\br,t)=J_0(k_0 r)\exp(-{r/\ell}),
  \end{equation}
 with $r\equiv |\br|$ and $J_0$ the Bessel function of the first kind. Eq. (\ref{g1_J0}) coincides with first-order correlation function of a 2D speckle pattern, which is here a ``matter-wave'' speckle formed by the particles scattered on the random potential \cite{Cherroret08}. The coherence length is very short in the multiple scattering regime, of the order of the de Broglie wavelength $2\pi/k_0$. In the presence of weak interactions and as long as $t\ll\tau_\text{col}$, the diffusive ring (of radial width $1/\ell$) broadens, while the energy distribution remains approximately centered on the initial kinetic energy $k_0^2/2m$. The argument of the exponential envelope in Eq. (\ref{g1_nointeractions}) therefore increases slowly in time while the $J_0$ function is hardly modified, which  scrambles the spatial oscillations of $g_1$ \cite{Scoquart21}.
 
A completely different physics shows up in the opposite limit $g\rho_0>V_0>\epsilon_0$ \cite{Scoquart21}. Since the interaction energy is now the largest energy scale, one expects the Bose gas to behave more like a \emph{superfluid}. This implies, in particular, that multiple scattering ceases and that the gas acquires a long-range spatial coherence. A manifestation of this coherence is visible in the momentum distribution shown in the inset of Fig. \ref{g1_SFregime}, here obtained for $\epsilon_0=0$:
\begin{figure}
\centering\includegraphics[scale=0.43]{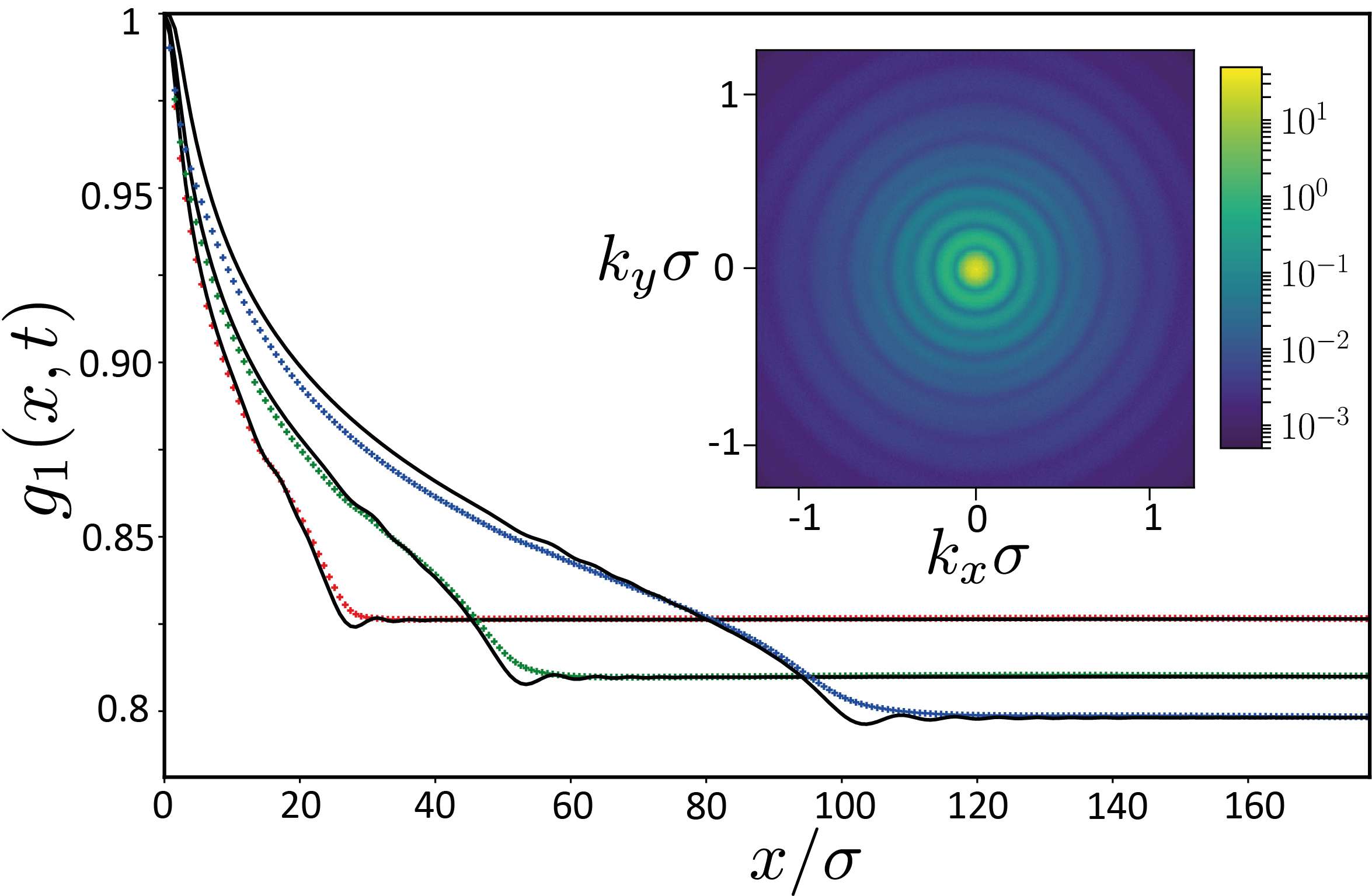}
\caption{
\label{g1_SFregime}
Cut along $x$ of the coherence function $g_1(\br,t)$, computed numerically by letting a plane-wave state $|\bk_0=0\rangle$ evolve with the Gross-Pitaevskii equation in a speckle potential (see \cite{Scoquart21} for the detailed parameters).  
Here $g\rho_0/V_0=3.75>1$ ($V_0$ is the standard deviation of the random potential), so that the disordered Bose gas is typically superfluid. Its correlations are long-range and spread within a light cone.
Symbols are the numerical results obtained  at times $t=15/(g\rho_0)$, $30/(g\rho_0)$ and $60/(g\rho_0)$ from bottom to top. 
Solid curves are the theoretical prediction (\ref{eq_g1_largeg}), using $\beta$ as a fit parameter for describing phonon collisions. The inset shows a density plot of the corresponding momentum distribution $n_\bk(t)$ for $t=30/g\rho_0$.
}
\end{figure}
in the superfluid regime, the disorder-induced field fluctuations are coherently enhanced and give rise to interference rings in momentum space. Such interference occurs because the Bose gas dynamics is now controlled by low-lying phonon excitations. These phonons both interfere with one another and with the Bose mean field. To see this quantitatively, one has to rely on the density-phase representation of the Bose gas, to be substituted for $\Psi$ in the Gross-Pitaevskii equation (\ref{eq:grosspitaevskii}):
\begin{equation}
\Psi(\br,t)=\sqrt{\rho(\br,t)}\exp[i\theta(\br,t)-ig\rho_0 t]/\sqrt{N}.
\end{equation}
In this representation the phase $-ig\rho_0 t$ describes the evolution of the uniform solution in the absence of disorder ($\rho(\br,t)=\rho_0$). In the presence of disorder, the phase $\theta$ and the density $\rho$ fluctuate. As long as $V_0\ll g\rho_0$ however, the density fluctuations remain small and one can resort to perturbation theory, writing $\rho(\br,t)=\rho_0+\delta\rho(\br,t)$ with $\delta\rho\ll\rho_0$. By linearizing the Gross-Pitaevskii equation, we then obtain a closed set of differential Bogoliubov-de-Gennes equations for $\theta$ and $\delta\rho$. Solving them together with the initial conditions $\rho(\br,0)=\rho_0$ and $\theta(\br,0)=0$ and averaging over disorder  eventually gives access to the  coherence function. In two dimensions one finds: 
\begin{align}
\label{eq_g1_largeg}
\ln g_1(\br,t)=-\!\!\int\!\!\frac{d^2\bk}{(2\pi)^2}B(\bk)(1-\cos\bk\cdot\br)\left(\left[\frac{1-\cos(E_k t)}{\epsilon_k+2g\rho_0} \right]^2\!\!+
\left[\frac{\sin(E_k t)}{E_k}\right]^2\!\!+\frac{\beta(t)}{k}\right),
\end{align}
where $B(\bk)\!\equiv\!\int d\br e^{i\bk\cdot \br}\overline{V(0)V(\br)}$
is the disorder power spectrum, $\epsilon_k=\bk^2/2m$ and $\smash{E_k=\sqrt{\epsilon_k(\epsilon_k+2g\rho_0)}}$ is the Bogoliubov dispersion relation.
The parameter $\beta(t)/k$ is a phenomenological correction term that takes into account corrections to the linearization approach, i.e. phonon collisions \cite{Scoquart21}.  Eq.  (\ref{eq_g1_largeg}) describes the dynamical spreading of correlations in a 2D Bose gas initially quenched from a weakly fluctuating state \cite{Berges04, Gring12, Cheneau12, Martone18}. 
In the present case however, the fluctuations are  are neither quantum or thermal but stem from the random potential. 
The prediction (\ref{eq_g1_largeg}) is compared with direct simulations of the Gross-Pitaveskii equation, using $\beta$ as a fitting parameter. The curves demonstrate the spreading of correlations within a light cone of transverse size $2c_s t$, where $c_s=\sqrt{g\rho_0/m}$ is the speed of sound. At large separations $r>2c_s t$, the correlation function displays a plateau reminiscent of the perfect coherence of the initial plane-wave state. Within the light cone on the other hand, the coherence function decays algebraically:
\begin{equation}
g_1(\br,t)\propto \left(\dfrac{\xi}{r}\right)^{\alpha}
\label{g1_algebraic}
\end{equation}
where $\xi\equiv 1/\sqrt{g\rho_0m}$ is the healing length and $\alpha\equiv(V_0\sigma/\sqrt{2}g\rho_0\xi)^2$. Interestingly, the algebraic law (\ref{g1_algebraic}) is independent of the precise shape of the disorder spectrum $B(\bk)$ and is thus, to a large extent, universal (it nevertheless depends on the disorder correlation length $\sigma$). Furthermore, the interior of the light cone varies very slowly in time. These properties are characteristic of a \emph{pre-thermalization} process of the Bose gas \cite{Berges04, Gring12, Cheneau12}, which here shows up in a disordered potential.  The long-range coherence within the light cone reflects the superfluid character of the gas, in strong contrast with the multiple scattering regime discussed in Sec. \ref{Sec_mesoscopic_regime} (compare with Eq. (\ref{g1_nointeractions})).

\section{Conclusion}

We have reviewed, in a cold-atom context, a few emblematic mesocopic phenomena that show up when a Bose matter wave dynamically evolves in a random potential: coherent backscattering, coherent forward scattering and the mesoscopic echo. While some of these phenomena were already known in condensed-matter and optical systems, their experimental observation in quantum gases is more recent and was achieved with an unprecedented control thanks to the versatility of cold-atom setups. Our understanding of the interplay between disorder and interactions in out-of-equilibrium configurations remains, on the other hand, in its early stages. In this paper we have addressed this question for bosons at the mean-field level. While in this limit long-time thermalization seems to be the rule whatever the disorder strength, the results presented in this paper show that the transient dynamics leading to this final state can be non-trivial. For a quenched  Bose gas initially prepared in a uniform state,  in particular, we have identified two well distinct non-equilibrium scenarios. A mesoscopic regime at weak enough interaction strength, where particle collisions reduce to an effective dephasing mechanism for weak localization, and a superfluid regime where multiple scattering is absent and the disordered Bose gas displays long-range spatial correlations spreading within a light cone. 
Even though strong enough disorder in one dimension is known to give rise to many-body localization and to suppress thermalization at the many-body level, a full picture of the non-equilibrium dynamics of disordered of quantum gases is still missing. The cross-over from the mean-field to the many-body level or the dynamics of quantum gases in dimensions larger than one remain, in particular, major challenges.

\section*{Acknowledgments}
Nicolas Cherroret and Dominique Delande acknowledge  financial support from the Agence Nationale de la Recherche (grants ANR-19-CE30-0028-01 CONFOCAL and ANR-18-CE30-0017 MANYLOK).


\begin{thebibliography}{99}
\bibitem{footnote}
As shown in \cite{Scoquart20}, interaction diagrams involving a a single particle collision in between any two scattering event on the random potential boil down to a mere energy shift $2g\rho_0$ of the energy distribution, and thus do not play any role in the dynamics.
\end{thebibliography}

\end{document}